%% file: main.tex
\documentclass[twocolumn,letterpaper,aps,prc,superscriptaddress,showpacs,floatfix,preprintnumbers,longbibliography]{revtex4-1}

\usepackage{graphicx}
\usepackage{amsmath}
\usepackage{hyperref}
\begin{document}


\title{Energy Independence of the Collins Asymmetry in $p^{\uparrow}p$ Collisions}

\input{star-author-list-2025-11-18.listB_aps.txt}

\date{\today}

\begin{abstract} 
The STAR experiment reports new, high-precision measurements of the transverse single-spin asymmetries for $\pi^{\pm}$ within jets, namely the Collins asymmetries, from transversely polarized ${p^{\uparrow}p}$ collisions at $\sqrt{s}$ = 510 GeV. 
The energy-scaled distribution of jet transverse momentum, $x_{\mathrm{T}} = 2p_{\mathrm{T,jet}}/\sqrt s$, shows a remarkable consistency for Collins asymmetries of $\pi^{\pm}$ in jets between $\sqrt{s}$ = 200 GeV and 510 GeV. This indicates that the Collins asymmetries are nearly energy independent with, at most, a very weak scale dependence in $p^{\uparrow}p$ collisions.
These results extend to high-momentum scales ($Q^2 \leq 3400$ GeV$^2$) and enable unique tests of evolution and universality in the transverse-momentum-dependent formalism, thus providing important constraints for the Collins fragmentation functions.
\end{abstract}

\maketitle



The very nature of the strong interaction obscures the mechanisms that drive the formation of the colorless hadrons from the fragmentation of high-energy quarks and gluons. A natural way to probe the process of hadronization is to study the spatial and momentum distributions of final-state hadrons and their dependence on the momentum, spin and flavor of the parent parton. Significant correlations have been observed, for example in the azimuthal distributions of hadrons produced in transversely polarized hadron-hadron collisions since 1970s~\cite{AN1977,E7042,2008AN,STARCollins2018,Rhicf,PHENIX:2020mft,STAR:2020nnl,STAR:2022hqg}. Initially it was a challenge for collinear perturbative QCD to describe the magnitude and momentum dependence of these transverse single-spin asymmetries (TSSAs)~\cite{Kane:1978nd}. This tension spawned the development of two QCD-based frameworks~\cite{DAlesio:2007bjf,Chen:2015tca} .
One is based on transverse-momentum-dependent (TMD) parton distribution functions (PDFs)~\cite{Sivers} or fragmentation functions
(FFs)~\cite{Collins:1992kk,Collins:1993kq}, and the other is on high twist PDF or FF under twist-3 collinear factorization~\cite{twist3_1,twist3_2,twist3_3}. 
The high twist framework applies to measurements with a single hard scale,
while the TMD framework applies to processes where there is another soft scale in addition to the hard scale, which can be as small as $\Lambda_{\mathrm{QCD}}$.
It has been shown that these approaches describe the same physics in the kinematic region where they overlap~\cite{TMDunite,Yuan:2009dw}. 
The high twist formalism attributes TSSA to multi-parton correlation functions~\cite{twist3_1,twist3_2,twist3_3}. 
In the TMD approach, TSSAs arise from the TMD PDF or FF, most notably the Sivers~\cite{Sivers} and Collins functions~\cite{Collins:1992kk,Collins:1993kq}, which encapsulate initial- and final-state spin-momentum correlations within a factorized description.
%
TMD functions play a crucial role in advancing our understanding of the three-dimensional structure of the nucleon in momentum space. 

The Collins fragmentation function, one of the key TMD FFs,  
describes the azimuthal distribution of hadrons produced from a fragmenting transversely polarized quark. An important process to probe Collins fragmentation functions is the study of the azimuthal asymmetries of hadrons inside jets, i.e.~Collins asymmetry, from transversely polarized proton–proton collisions~\cite{FengCollins2008}, which are only available at the Relativistic Heavy Ion Collider (RHIC)~\cite{Hahn:2003sc}. 
The Collins asymmetry in $p^{\uparrow}p$ collisions arises from the convolution of the collinear quark transversity with the TMD Collins FF~\cite{FengCollins2008,Kang:2017btw, Kang:2017glf}, $H^{\perp}_{1\,\pi/q}\left(z,j_{\mathrm{T}},Q\right)$. Here, $Q$ is the TMD factorization scale, $z$ is the hadron momentum fraction within the jet, and $j_{\mathrm{T}}$ is the hadron momentum transverse to the jet axis. 
The transversity distribution describes the transverse polarization
of quarks in a transversely polarized proton. Study of the Collins FF provides an excellent opportunity to investigate fundamental QCD properties such as factorization, universality, and evolution of TMDs. 
It has been shown that the Collins FF measured for hadrons within reconstructed jets in $p^{\uparrow}p$ collisions is universal~\cite{FengCollins2008,Kang:2017btw}, \textit{i.e.}, the same as in semi-inclusive deep-inelastic scattering (SIDIS) and $e^{+}e^{-}$ collisions. 
The $Q$-dependence is generally referred to as TMD evolution.  It must be inferred from experiment because it includes a contribution that cannot be calculated perturbatively~\cite{Echevarria:2014xaa,Kang:2015msa}.
One has to introduce a prescription for the non-perturbative part and
a moderate but visible suppression of the Collins asymmetry has been predicted with the growth of $Q^2$ in SIDIS and $pp$ collisions~\cite{Kang:2015msa,Kang:2017btw,Zeng:2024gun}. In contrast, a significant suppression of the Sivers asymmetry in W/Z 
production in $pp$ collisions has also been predicted~\cite{Echevarria:2014xaa,Bury:2021sue,Bury:2020vhj,Bacchetta:2020gko,STAR:2015vmv,STAR:2023jwh}.  

Recently, the STAR experiment~\cite{STAR:2002eio} at RHIC  made the first and, to date, only observations of the Collins asymmetry of $\pi^\pm$ in jets in $p^{\uparrow}p$ collisions. This has been achieved at $\sqrt{s}$ = 200 GeV with high statistics~\cite{STAR:2022hqg}, and also at 500 GeV with a smaller data sample~\cite{STARCollins2018}. 
These measurements provide similar coverage over the range of incident parton longitudinal momentum fraction $x$ as the current SIDIS measurements~\cite{SIDIS1,SIDIS2,SIDIS3,SIDIS4,SIDIS5,SIDIS6,SIDIS7}, but at much higher values of momentum transfer $Q^2$. This positions STAR as an ideal setting to test the evolution of TMD functions.
The STAR Collins asymmetry measurements at $\sqrt s$ = 200 and 500 GeV at mid-rapidity have been compared over their common region of $x_{\mathrm{T}}=2p_{\mathrm{T,jet}}/\sqrt{s}$, 
thus allowing for comparisons of measurements at similar momentum fraction regions ($x \simeq x_{\mathrm{T}}e^{\pm \eta}$). 
The results were statistically consistent, suggesting that the Collins effect has a weak energy dependence in hadronic collisions~\cite{STAR:2022hqg}. However, the comparatively small size of the existing 500 GeV data sample~\cite{STARCollins2018} limited the conclusions that could currently be drawn, thereby preventing any determination of TMD-evolution effects.

In this letter, the STAR Collaboration presents high-precision measurements of the Collins asymmetry for $\pi^\pm$ in jets from transversely polarized $p^{\uparrow}p$ collisions at $\sqrt s = 510$ GeV, based on a dataset with 13 times the luminosity of the previously published 500 GeV results~\cite{STARCollins2018}. Comparing these findings with the 200 GeV results~\cite{STAR:2022hqg} places strong constraints on the TMD evolution of the Collins function. 
Currently, our understanding of TMD evolution is far less complete than that of collinear evolution, and experimental measurements of Collins asymmetry in $pp$ processes at different energies have also been rather scarce.
Additionally, the results are compared with theoretical predictions using quark transversity and Collins fragmentation functions derived from SIDIS and electron-positron annihilation, providing an important test of the universality of the Collins fragmentation function across different processes.


The data were collected by the STAR experiment in 2017 from transversely polarized $p^{\uparrow}p$ collisions at $\sqrt s$ = 510 GeV. The integrated luminosity was about 320 pb$^{-1}$ with an average beam polarization of 55.0 $\pm$ 1.4\%~\cite{ref:RHICPolG}. The detector subsystems utilized for this analysis include the time projection chamber (TPC)~\cite{TPC}, the barrel~\cite{BEMC} and endcap (BEMC and EEMC)~\cite{EEMC} electromagnetic calorimeters, and the time-of-flight (TOF)~\cite{TOF,Chen:2024aom} detector. Events were selected using jet patch (JP) triggers, which required the transverse electromagnetic energy within a $1.0 \times 1.0$ region in pseudorapidity ($\eta$) $\times$ azimuth ($\phi$) space in the BEMC and EEMC to exceed one of several predefined energy thresholds.

The analysis procedures closely follow those of the previous STAR measurements at 200 GeV~\cite{STAR:2022hqg} and 500 GeV~\cite{STARCollins2018}. Jets are reconstructed using the anti-$k_T$ algorithm~\cite{antikT} with a resolution parameter $R$ = 0.5 to reduce the contributions from soft background.  Collins asymmetries were calculated for hadrons with $0.1 < z < 0.8$ within jets spanning $0 < \eta < 0.9$ relative to the polarized beam. 
To further limit backgrounds and minimize the underlying-event-subtraction uncertainty, an upper limit on $j_{\mathrm{T}}$ was enforced as in \cite{STAR:2022hqg}, which was optimized with respect to $p_{\mathrm{T,jet}}$ and hadron $z$ to exclude hadrons within the kinematic region dominated by the underlying event. 
A similar requirement, \( j_{\mathrm{T}} \le j_{\mathrm{T}, \text{max}} = \text{min}[((0.0093 + 0.2982 \cdot z) \cdot p_{\mathrm{T,jet}}), 2.5~\text{GeV}/c] \), was imposed for hadrons to be included in the Collins asymmetry calculations in this analysis. 
A minimum distance between the hadron and the jet axis, \( \Delta R_h = \sqrt{(\Delta \eta)^2 + (\Delta \phi)^2} \), was set in the previous measurements~\cite{STARCollins2018,STAR:2022hqg} to ensure a robust reconstruction of the Collins angles. This cut is tightly correlated both to hadron $z$ and $j_{\mathrm{T}}$ and can be approximated by $j_{\mathrm{T,min}}  \approx z \times \Delta R_\mathrm{h,min} \times p_{\mathrm{T,jet}}$.  It was set to 0.05 in the 200 GeV analysis. 
To align the acceptance in the current analysis at a given $z$ and $x_{\mathrm{T}}$ with the 200 GeV measurement, this minimum distance was reduced to 0.02.  This necessitated a finite angular resolution correction to the measured asymmetries of 25\% for the lowest $p_{\mathrm{T,jet}}$ jets, dropping to 10\% at $p_{\mathrm{T,jet}} = 14$ GeV/$c$, and $\le$\,5\% at $p_{\mathrm{T,jet}} > 30$ GeV/$c$.

Charged pion candidates were selected if their ionization energy loss ($dE/dx_{\mathrm{meas}}$) in the TPC was consistent with the calculated value for pions of the measured momentum ($dE/dx_{\mathrm{\pi,calc}}$). The difference was normalized with the experimental resolution ($\sigma_{dE/dx}$)~\cite{Shao:2005iu} to obtain:
\begin{equation}
n_\sigma(\pi)=\frac{1}{\sigma_{dE/dx}} \ln \left(\frac{dE/dx_{\mathrm{meas}}}{dE/dx_{\mathrm{\pi,calc}}}\right) .
\end{equation}
Along with the TPC, TOF was used to improve the pion purity for momenta up to approximately 3 GeV/$c$ by measuring the particles' arrival times. The TOF measurement was normalized to 
\begin{equation}
n_{\sigma,\mathrm{TOF}}(\pi)=\frac{\mathrm{TOF}_{\mathrm{meas}}-\frac{L}{ \mathrm{c} \beta_{\pi}(p)}}{\sigma_{\mathrm{TOF}}} ,
\end{equation}
where $\mathrm{TOF}_{\mathrm{meas}}$ is the measured flight time of the particle, $L$ is the path length, $c$ is the speed of light, $\beta_{\pi}(p)$ is the velocity of pions at momentum $p$, and $\sigma_{\mathrm{TOF}}$ represents the resolution of the TOF time measurement. To identify the pions, $n_\sigma(\pi)$ was constrained to lie in the range \((-1, 2)\). In regions where TOF is effective, $n_{\sigma,\mathrm{TOF}}(\pi)$ was additionally limited to \( (-5, 3) \), further enhancing pion purity.\par

Simulated events are used to calculate corrections to the jet kinematics and to estimate the associated systematic uncertainties. They were generated using \textsc{Pythia} 6.4.28~\cite{ref:Pythia6} with the STAR-adjusted Perugia 2012 tune~\cite{ref:PerugiaTunes} (PARP(90) adjusted from 0.24 to 0.213~\cite{ref:STAR_2012_ALL}). PYTHIA events were processed through full detector simulations in \textsc{Geant}~3~\cite{ref:Geant}. These events were then embedded into randomly selected bunch crossing events collected during the 2017 510 GeV run to account for event pileup and beam background effects. The $p_{\mathrm{T,jet}}$, pion $j_{\mathrm{T}}$, and pion $z$ measured at the ``detector level" were each corrected back to the ``particle level." 
Here, the detector-level jets in simulation were reconstructed in the same way as in the data, using charged-particle tracks and calorimeter towers, while the particle-level jets were formed from the stable (as defined in~\cite{ref:PerugiaTunes}) final-state particles produced in the simulated collisions.\par

\begin{figure}[!hbt]
	\centering
	\includegraphics[width=0.85\linewidth]{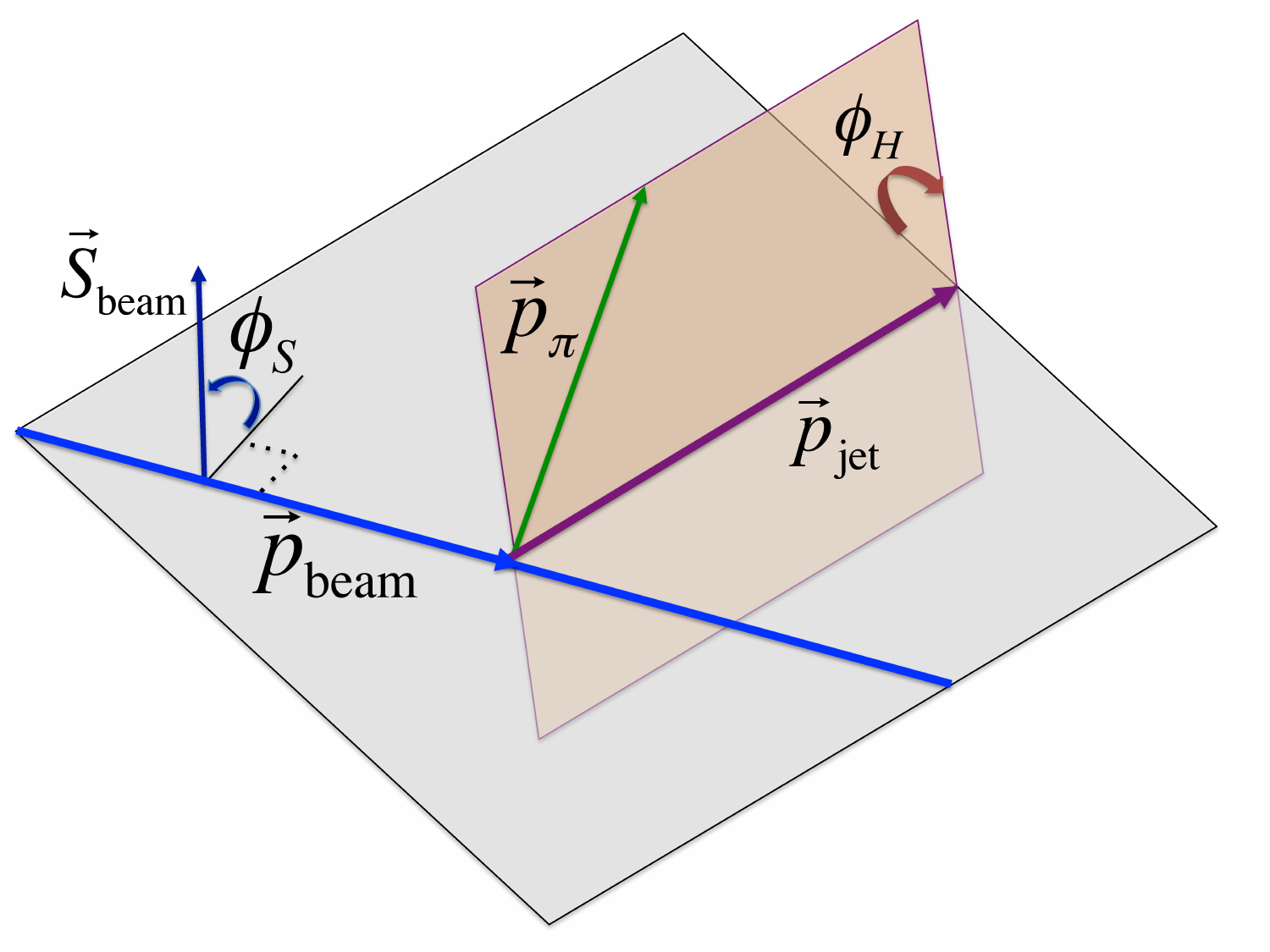}
	\caption{The representation of the jet scattering plane along with the definitions of the azimuthal angles $\phi_S$ and $\phi_H$~\cite{STARCollins2018}.}
	\label{fig:Phi_H}
\end{figure}

In transversely polarized $p^{\uparrow}p$ collisions, the Collins asymmetry of hadrons inside jets can be extracted through the spin-dependent azimuthal modulation of hadrons, using the cross-ratio method,
\begin{widetext}
\begin{equation}
\label{eq:asy8}
A_{\mathrm{UT}}^{\sin\phi_C}\sin \phi_C  \\
=\frac{1}{P}
\frac{\sqrt{N^{\uparrow}[\phi_S-\phi_H]~N^{\downarrow}[(\phi_S+\pi)-\phi_H]}-\sqrt{N^{\downarrow}[\phi_S-\phi_H]~N^{\uparrow}[(\phi_S+\pi)-\phi_H]}}{\sqrt{N^{\uparrow}[\phi_S-\phi_H]~N^{\downarrow}[(\phi_S+\pi)-\phi_H]}+\sqrt{N^{\downarrow}[\phi_S-\phi_H]~N^{\uparrow}[(\phi_S+\pi)-\phi_H]}} ,
\end{equation}
\end{widetext}
which accounts for detector efficiencies and spin-dependent luminosities~\cite{cross-ratio}.  $N^{\uparrow}$ ($N^{\downarrow}$) is the hadron yield when the beam spin is ``up" (``down"), and $P$ is the beam polarization. The azimuthal moment of the Collins asymmetry is isolated by the $\sin\phi_C = \sin(\phi_S - \phi_H)$ dependence, where $\phi_S$ is the angle between the proton polarization direction and the jet scattering plane, and $\phi_H$ is the pion’s transverse azimuthal angle relative to the jet axis, as illustrated in Fig.~\ref{fig:Phi_H}.  The jet scattering plane is defined by the incoming polarized proton beam and scattered jet momentum. In Eq.~(\ref{eq:asy8}), $\phi_S$ spans the range $-\pi/2$ to $\pi/2$, and $\phi_H$ spans the full azimuth.


To exclude contamination from kaons, protons, and electrons, similar procedures as in the previous measurements were applied to extract the pure pion asymmetries~\cite{STARCollins2018,STAR:2022hqg}. In this approach, $\pi/K/p/e$-rich samples were first identified using the TPC and TOF information. Raw asymmetries, which represent linear mixtures of the pure asymmetries of each particle type, were calculated for each enriched sample. Using particle fractions determined from TOF and TPC data, the pure pion asymmetries were then extracted.


Particle identification and corresponding particle fractions $f_{i_{rich}}^j$, which represent the fraction of particle type $j$ in the $i$-rich sample, were obtained using the $dE/dx$ from the TPC, and flight time from TOF when matching to available TOF hits, given detector acceptance and efficiency limitations.
About half of the TPC tracks have TOF information.
Raw asymmetries and particle fractions were determined separately in each particle-rich region, with and without TOF data. The relationship between the raw and pure asymmetries is:
\begin{equation}
\label{eq:asy5}
\resizebox{0.90\linewidth}{!}{$
\begin{pmatrix}
f^{\pi_\mathrm{TOF}}_{\pi_\mathrm{rich}} & f^{K_\mathrm{TOF}}_{\pi_\mathrm{rich}} & f^{p_\mathrm{TOF}}_{\pi_\mathrm{rich}} \\
f^{\pi_\mathrm{TOF}}_{K_\mathrm{rich}} & f^{K_\mathrm{TOF}}_{K_\mathrm{rich}} & f^{p_\mathrm{TOF}}_{K_\mathrm{rich}} \\
f^{\pi_\mathrm{TOF}}_{p_\mathrm{rich}} & f^{K_\mathrm{TOF}}_{p_\mathrm{rich}} & f^{p_\mathrm{TOF}}_{p_\mathrm{rich}} \\
f^{\pi_{dE/dx}}_{\pi_\mathrm{rich}} & f^{K_{dE/dx}}_{\pi_\mathrm{rich}} & f^{p_{dE/dx}}_{\pi_\mathrm{rich}} \\
f^{\pi_{dE/dx}}_{K_\mathrm{rich}} & f^{K_{dE/dx}}_{K_\mathrm{rich}} & f^{p_{dE/dx}}_{K_\mathrm{rich}} \\
f^{\pi_{dE/dx}}_{p_\mathrm{rich}} & f^{K_{dE/dx}}_{p_\mathrm{rich}} & f^{p_{dE/dx}}_{p_\mathrm{rich}} \\
\end{pmatrix}
\begin{pmatrix}
A^{\mathrm{pure}}_{\mathrm{UT}, \pi} \\
A^{\mathrm{pure}}_{\mathrm{UT}, K} \\
A^{\mathrm{pure}}_{\mathrm{UT}, p}
\end{pmatrix}
=
\begin{pmatrix}
A_{\mathrm{UT}, \pi_\mathrm{rich}}^\mathrm{TOF} \\
A_{\mathrm{UT}, K_\mathrm{rich}}^\mathrm{TOF} \\
A_{\mathrm{UT}, p_\mathrm{rich}}^\mathrm{TOF} \\
A_{\mathrm{UT}, \pi_\mathrm{rich}}^{dE/dx} \\
A_{\mathrm{UT}, K_\mathrm{rich}}^{dE/dx} \\
A_{\mathrm{UT}, p_\mathrm{rich}}^{dE/dx} \\
\end{pmatrix}
$}
\end{equation}

Electron fractions do not appear because $A_{\mathrm{UT},e}$ is expected~\cite{STAR:2022hqg} -- and was 
found -- to be zero.
The pure asymmetries were then extracted by $\chi^2$ minimization of Eq.~(\ref{eq:asy5}) utilizing Moore-Penrose pseudoinversion~\cite{matrix1,matrix2}.

\begin{figure}[!hbt]
	\centering
	\includegraphics[width=0.99\linewidth]{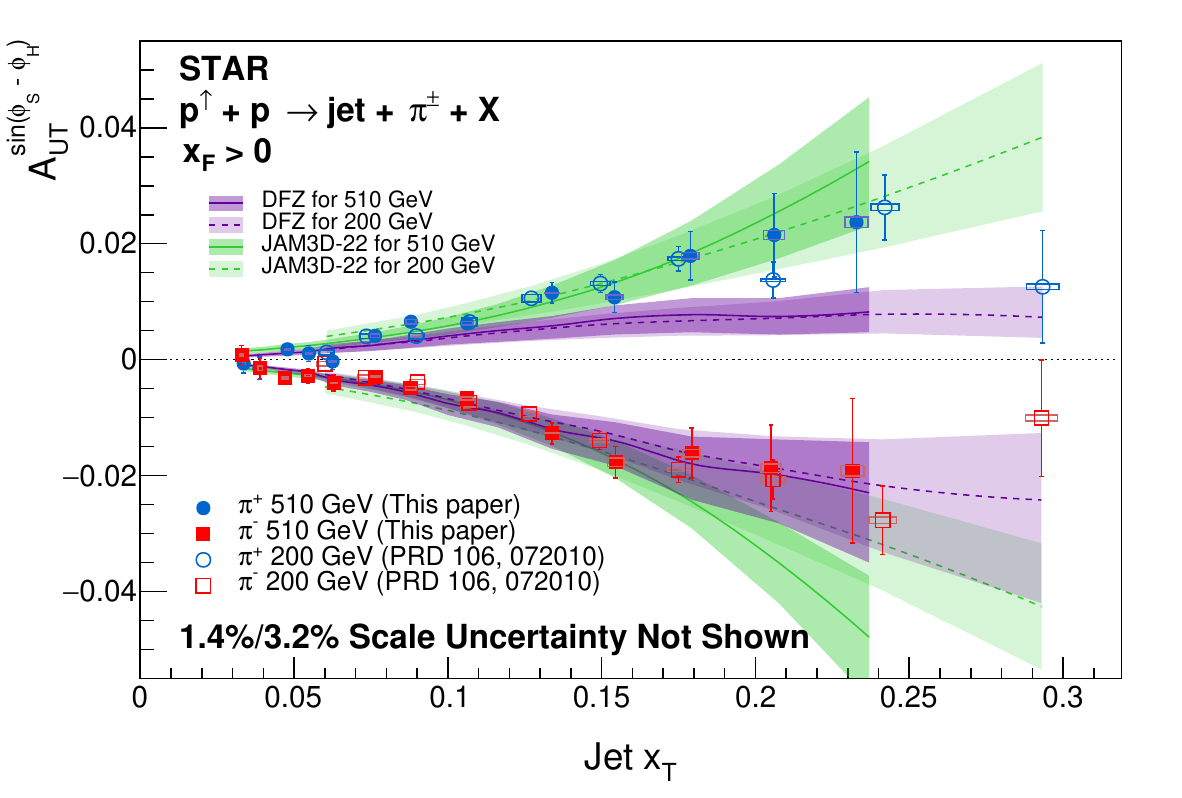}
	\caption{Collins asymmetries, $A_{\mathrm{UT}}^{\sin(\phi_S - \phi_H)}$, as a function of jet $x_{\mathrm{T}}$ ($\equiv \frac{2p_{\mathrm{T,jet}}}{\sqrt{s}}$) for $\pi^{\pm}$ in $p^{\uparrow}p$ collisions at $\sqrt{s} = 510$ GeV (solid points), compared with previous results at $\sqrt{s} = 200$ GeV (open points). Vertical bars show the statistical uncertainties; boxes show the systematic uncertainties in $x_{\mathrm{T}}$ and $A_{\mathrm{UT}}$.}
	\label{fig:AUT_xT}
\end{figure}

Figure~\ref{fig:AUT_xT} shows the Collins asymmetries for $\pi^\pm$ in jets at $\sqrt s$ = 510 GeV (solid points) as a function of the jet $x_{\mathrm{T}}$, compared with previously published results at 200 GeV (open points)~\cite{STAR:2022hqg}. 
The primary source of systematic uncertainty in $A_{\mathrm{UT}}$, shown by the height of the uncertainty boxes, is trigger bias, which can reach up to 20\% in certain kinematic regions. This uncertainty was estimated based on simulations, as the STAR JP trigger system may have different efficiencies for quark- and gluon-initiated jets, thereby distorting the measured asymmetries.
 An overall systematic uncertainty of 3.2\% for the 200 GeV data and 1.4\% for the 510 GeV data, arising from beam polarization, is not displayed. Systematic uncertainties are discussed further in~\cite{supply}.
The dominant uncertainty in $x_{\mathrm{T}}$, shown by the width of the uncertainty boxes, arises from jet-energy-scale uncertainties, primarily due to the efficiency and resolution of the electromagnetic calorimeters.\par

The asymmetry clearly increases with jet $x_{\mathrm{T}}$, consistent with the expectation that magnitude of quark transversity grows with parton momentum fraction $x$.
As the asymmetries here are expected to originate from the convolution of quark transversity with the Collins function, the comparison of Collins asymmetries from two energies at a fixed $x_{\mathrm{T}}$ (corresponding to a similar $x$) reflects the effect of TMD evolution.
With high statistics of this analysis, a detailed comparison across different collision energies is now possible. As shown in the figure, the high-precision Collins asymmetry results at 510 GeV and 200 GeV align well within the overlap region $0.06 < x_{\mathrm{T}} < 0.25$, despite the corresponding \(Q^2\) values differing by a factor of about 6. 
To quantify the consistency of the Collins asymmetries between the two energies, a $t$-test was conducted that shows no statistically significant evidence of a difference in the asymmetries~\cite{supply}.
This strongly suggests a surprisingly weak TMD evolution for the Collins function in hadronic interactions. 
Our measurements are compared with two theoretical calculations, JAM3D-22~\cite{Gamberg:2022kdb} and DFZ~\cite{DAlesio:2025jmr}.
The JAM3D-22 analysis simultaneously includes TMD and collinear twist-3 observables without TMD evolution, based on global analysis of $e^+e^-$, SIDIS and $p^{\uparrow}p$ data~\cite{Gamberg:2022kdb}.
The DFZ calculation~\cite{DAlesio:2025jmr} employs a simplified TMD approach at leading order and a collinear configuration for the initial state, based on global fit of $e^+e^-$, SIDIS data without TMD evolution~\cite{Boglione:2024dal}. 
Neither of the two models includes the previous STAR Collins results in Ref.~\cite{STARCollins2018,STAR:2022hqg}, the data agree well with both models with $x_{\mathrm{T}}$ up to $\sim$0.2, consistent with theoretical models defined by weak energy dependence. 
The agreement with DFZ calculation supports the
universality of Collins function and TMD factorization. 

%
%



\begin{figure}[!hbt]
	\centering
	\includegraphics[width=0.99\linewidth]{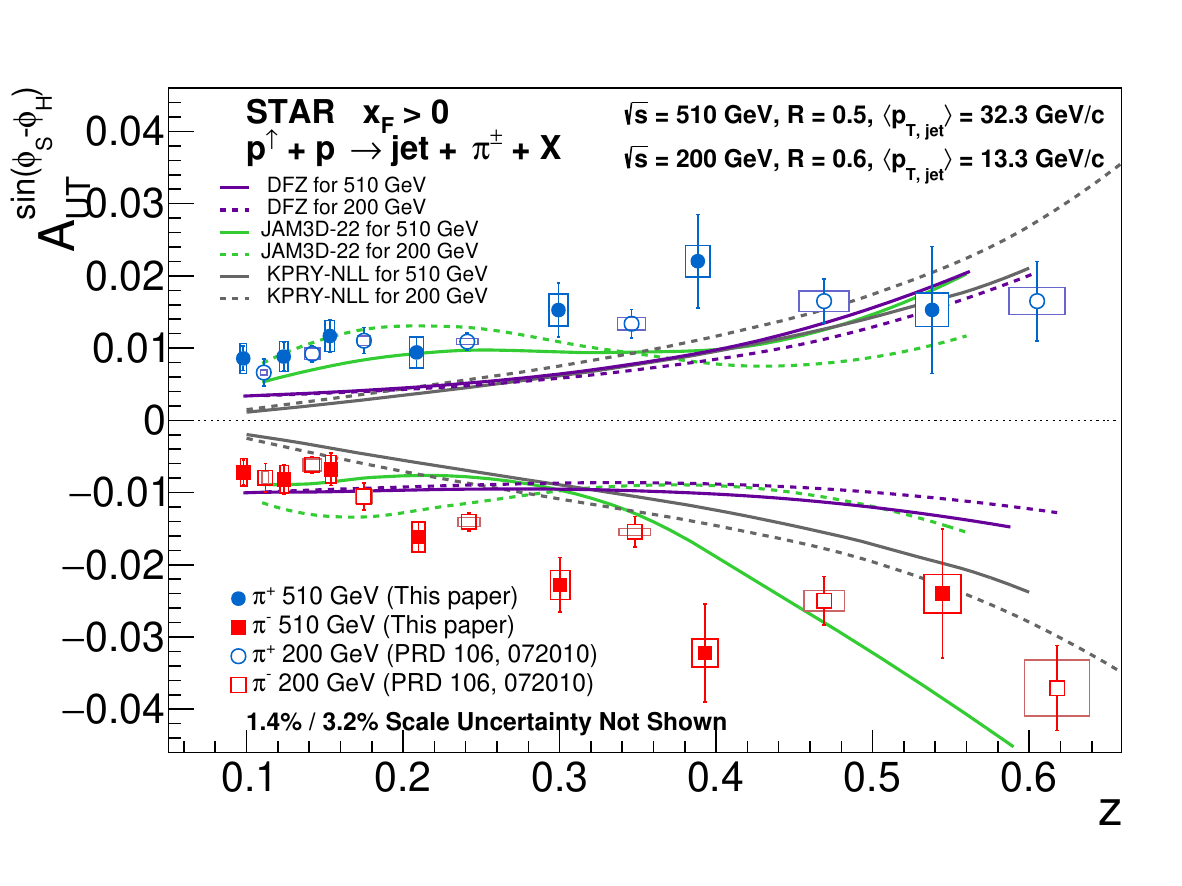}
	\caption{Collins asymmetries, $A_{\mathrm{UT}}^{\sin(\phi_S - \phi_H)}$ as a function of $\pi^{\pm}$ longitudinal momentum fraction $z$ in $p^{\uparrow}p$ collisions at $\sqrt{s} = 510$ GeV and 200 GeV. Vertical bars show the statistical uncertainties; boxes show the systematic uncertainties. Model uncertainties can be found in~\cite{supply}.} 
	\label{fig:AUT_z}
\end{figure}
Figure \ref{fig:AUT_z} shows the Collins asymmetries as a function of pion $z$. For the 510 GeV data, the $p_{\mathrm{T,jet}}$ range is integrated over $24.5 <p_{\mathrm{T,jet}}< 62.8$ GeV/$c$, to achieve comparable jet $x_{\mathrm{T}}$ values to the 200 GeV measurement with $p_{\mathrm{T,jet}}> 9.9$ GeV/$c$.
The asymmetries increase slowly with $z$, and again the results at the two energies are in good agreement under $t$-tests~\cite{supply}. 
The data are compared with three theoretical calculations KPRY~\cite{Kang:2017btw}, JAM3D-22 and DFZ.
The KPRY calculation uses a framework combining collinear and TMD factorization.
KPRY includes TMD evolution, extending to next-to-leading logarithmic accuracy, and is based on a global analysis of $e^+e^-$ and SIDIS data~\cite{Kang:2015msa}.
All three models follow the trend of our data in general, with weak energy dependence. 
Again, the comparison of our $p^{\uparrow}p$ data with the KPRY and DFZ calculation provides a test of universality of Collins function and TMD factorization in this process, since they only fits SIDIS and $e^+e^-$ data. 

\begin{figure}[!hbt]
	\centering
	\includegraphics[width=0.99\linewidth]{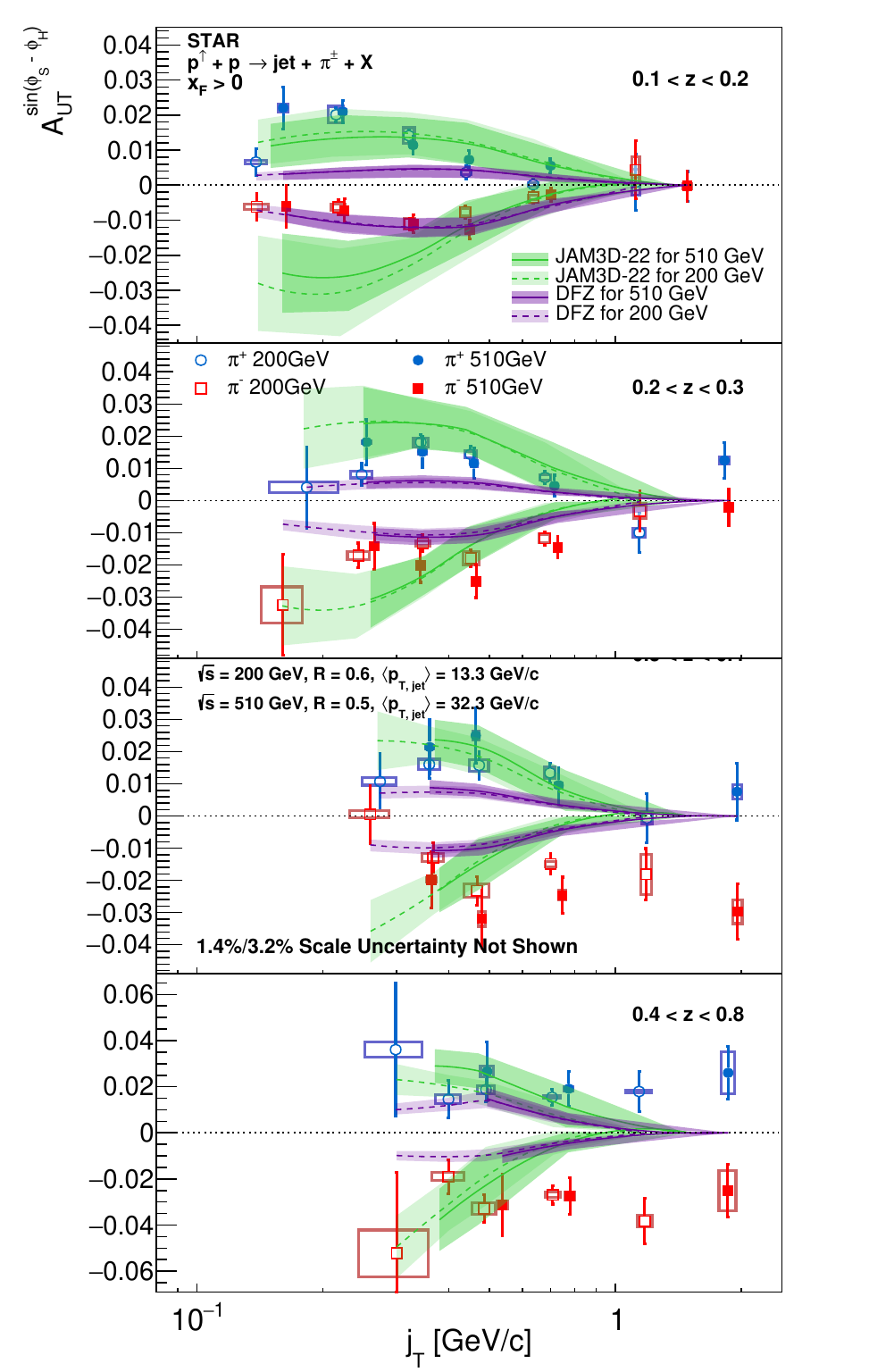 }
	\caption{Collins asymmetries, $A_{\mathrm{UT}}^{\sin(\phi_S - \phi_H)}$, as a function of the $\pi^{\pm}$ momentum transverse to the jet axis, $j_{\mathrm{T}}$, in four longitudinal momentum fraction $z$ bins. The solid points show the results from this analysis of $\sqrt{s} = 510$ GeV $p^{\uparrow}p$ collisions with an average $\langle p_{\mathrm{T,jet}} \rangle = 32.3$ GeV/c, while the open points show the results from $\sqrt{s} = 200$ GeV $p^{\uparrow}p$ collisions with an average $\langle p_{\mathrm{T,jet}} \rangle = 13.3$ GeV/c~\cite{STAR:2022hqg}. Vertical bars show the statistical uncertainties and boxes show the systematic uncertainties in $j_{\mathrm{T}}$ and $A_{\mathrm{UT}}$.}
	\label{fig:jt}
\end{figure}

Figure \ref{fig:jt} shows the Collins asymmetries as a function of pion $j_{\mathrm{T}}$ in four hadron-$z$ bins, with the same $p_{\mathrm{T,jet}}$ ranges as Fig. \ref{fig:AUT_z}. 
The $(j_{\mathrm{T}},z)$ dependence of the Collins asymmetries at the same $x_{\mathrm{T}}$ is consistent at the two beam energies, again indicating a very weak energy dependence of the Collins effect in hadronic collisions. These results are also compared with theoretical models, JAM3D-22 and DFZ. 
Both models generally follow the experimental data trend, yet significant discrepancies remain, in particular when $j_{\mathrm{T}}$ and $z$ are both large. 
Thus, these data offer crucial constraints on Collins fragmentation functions.

In summary, the STAR Collaboration presents high-precision measurements of the Collins asymmetries for $\pi^\pm$ inside jets from transversely polarized $p^{\uparrow}p$ collisions at $\sqrt s=510$ GeV. These new results reveal an excellent consistency with previous measurements at 200 GeV over the common $x_{\mathrm{T}}$ region, providing evidence for no significant energy evolution of the Collins asymmetry in hadronic interactions.
The results are in good agreement with theoretical models that do not implement TMD evolution. 
This apparent energy independence is in striking contrast to predictions that TMD evolution of the Sivers function leads to a substantial suppression of TSSA for $W^{\pm}$ production \cite{Echevarria:2014xaa,Bury:2021sue,Bury:2020vhj,Bacchetta:2020gko,STAR:2015vmv}.
\par

A good agreement between our $pp$ data and theoretical models (KPRY and DFZ) that only include $e^+e^-$ and SIDIS data, supports the universality of Collins function and TMD factorization in this process.
Discrepancies are still observed in the results versus $j_{\mathrm{T}}$, indicating the need for an improved theoretical understanding of the physics in this region. These results provide essential information on transverse spin-dependent effects, especially regarding quark transversity and Collins fragmentation functions. 
With upgraded detectors at forward rapidities, STAR collected additional datasets at both 200 GeV and 510 GeV, which represent the final opportunity in the near term to explore factorization, universality, and evolution of the Collins asymmetry. In the long term, these $pp$ data will provide the baseline for the ultimate universality test when compared to similar measurements in $ep$ collisions at the future Electron-Ion Collider (EIC).


Acknowledgment:
We thank Umberto D’Alesio, Carlo Flore, Marco Zaccheddu, Daniel Pitonyak, and Alexei Prokudin for valuable discussions and providing the results of their theoretical calculations. 
We thank the RHIC Operations Group and SDCC at BNL, the NERSC Center at LBNL, and the Open Science Grid consortium for providing resources and support.  This work was supported in part by the Office of Nuclear Physics within the U.S. DOE Office of Science, the U.S. National Science Foundation, National Natural Science Foundation of China, Chinese Academy of Science, the Ministry of Science and Technology of China and the Chinese Ministry of Education, NSTC Taipei, the National Research Foundation of Korea, Czech Science Foundation and Ministry of Education, Youth and Sports of the Czech Republic, Hungarian National Research, Development and Innovation Office, New National Excellency Programme of the Hungarian Ministry of Human Capacities, Department of Atomic Energy and Department of Science and Technology of the Government of India, the National Science Centre and WUT ID-UB of Poland, the Ministry of Science, German Bundesministerium f\"ur Bildung, Wissenschaft, Forschung and Technologie (BMBF), Helmholtz Association, Ministry of Education, Culture, Sports, Science, and Technology (MEXT), and Japan Society for the Promotion of Science (JSPS).

Data availability: The data that support the findings of this Letter are openly available~\cite{Data:availability}


\input{main.bbl}

\end{document}


\title{Supplemental material: Energy Independence of the Collins Asymmetry in $p^{\uparrow}p$ Collisions}
\author{The STAR Collaboration}
\date{\today}

\maketitle

\vspace*{2cm}

\mysection{PARTICLE IDENTIFICATION}

\begin{figure}[H]
    \centering
    \begin{subfigure}[b]{0.44\textwidth}
        \centering
        \includegraphics[width=\linewidth]{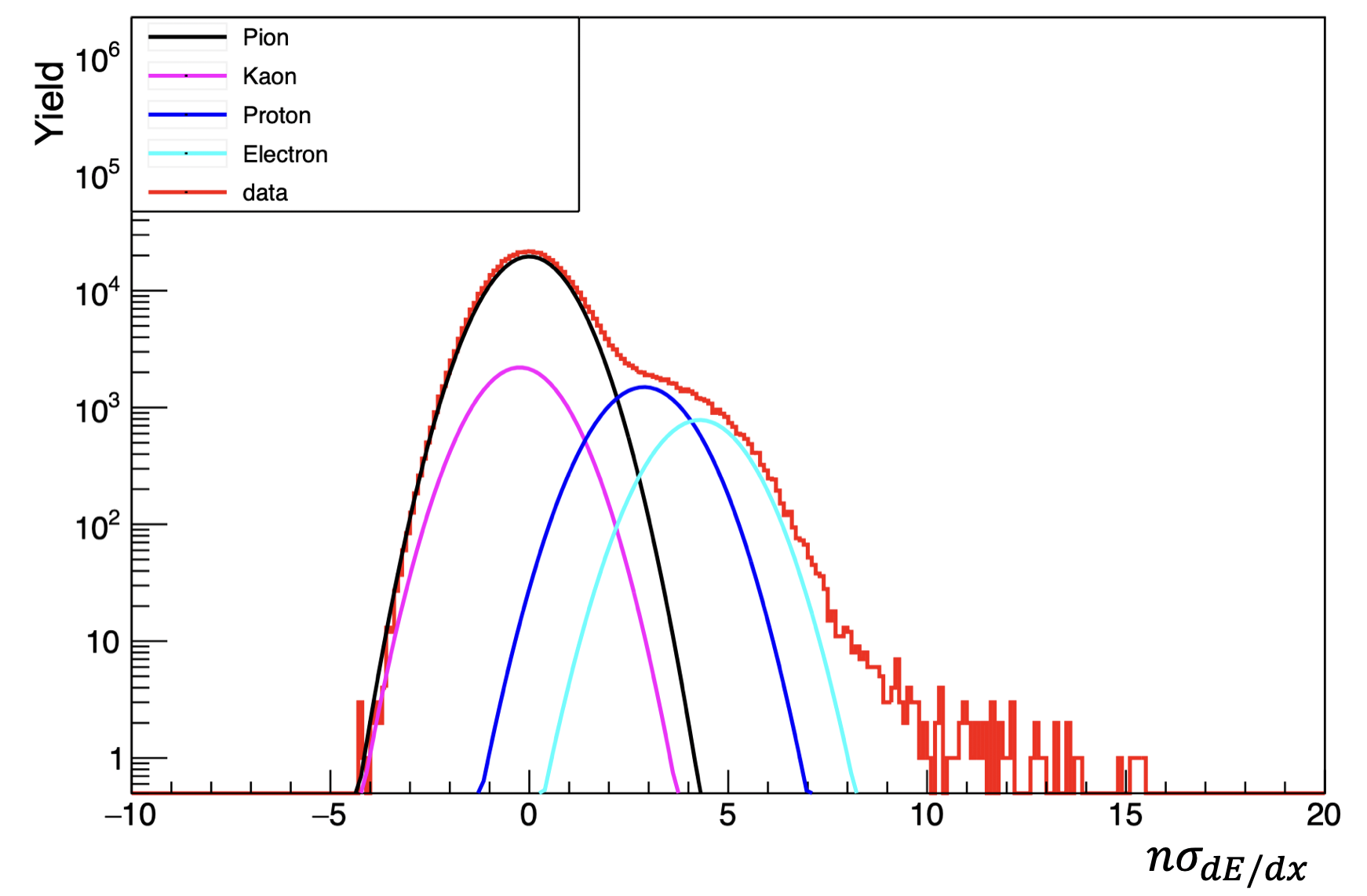}
        \caption{}
        \label{fig1:pid1}
    \end{subfigure}
    \hfill
    \begin{subfigure}[b]{0.48\textwidth}
        \centering
        \includegraphics[width=\linewidth]{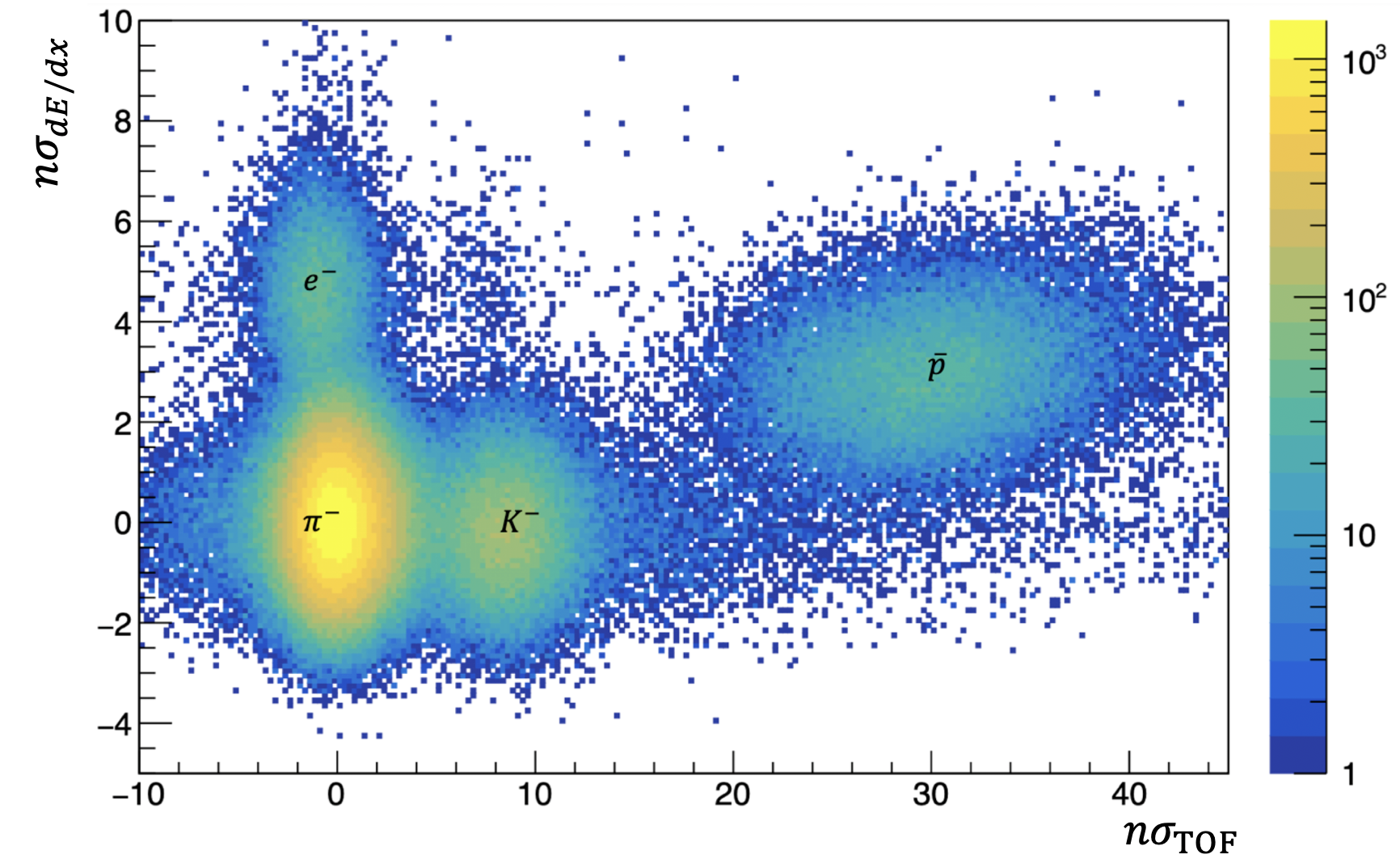}
        \caption{}
        \label{fig1:pid2}
    \end{subfigure}
    \caption{(a) $n\sigma_{dE/dx}$ distribution with multi-Gaussian fits. (b)  The correlations of $n\sigma_{dE/dx}$ vs $n\sigma_{\mathrm{TOF}}$ for negatively charged particles carrying momentum fractions of $0.1 < z < 0.8$ in jets with $6.5 < p_{\mathrm{T,jet}} < 62.8$ GeV$/c$.}
    \label{fig:PID}
\end{figure}

The Collins asymmetry is measured for identified hadrons within reconstructed jets, which necessitates reliable charged-particle identification. The particle identification (PID) strategy employed in this work follows established procedures from previous STAR analyses. At STAR, PID is primarily achieved using the Time Projection Chamber (TPC) and the Time-of-Flight (TOF) detector. Fig.~\ref{fig:PID}a presents an example of the \(n\sigma_{dE/dx}\) distribution within the momentum range 1.15 GeV/$c$ $<$ $p$ $<$ 1.20 GeV/$c$, together with the corresponding Gaussian fits. The figure corresponds to jets with 6.5 GeV/$c$ $<$ $p_T$ $<$ 62.8 GeV/$c$ and a longitudinal momentum fraction range of $0.1 < z < 0.8$. The relative amplitudes of the Gaussian components provide an estimate of the expected fraction of each particle species.
In 510 GeV $pp$ collisions, we find that approximately half of the tracks are associated with TOF information. To ensure sufficient statistics for stable fits, the TOF is applied only for tracks with momentum $p < 3$ GeV/$c$. When TOF information is available, a two-dimensional identification is performed to the joint distribution of \(n\sigma_{dE/dx}\) and \(n\sigma_{\mathrm{TOF}}\). Fig.~\ref{fig:PID}b shows the corresponding distribution for the same kinematic range, demonstrating the improved particle separation achieved by the combined use of TPC and TOF.

\FloatBarrier

\mysection{COLLINS ASYMMETRY OF $e^{\pm}$}

\begin{figure}[H]
    \centering
    \includegraphics[width=0.5\textwidth]{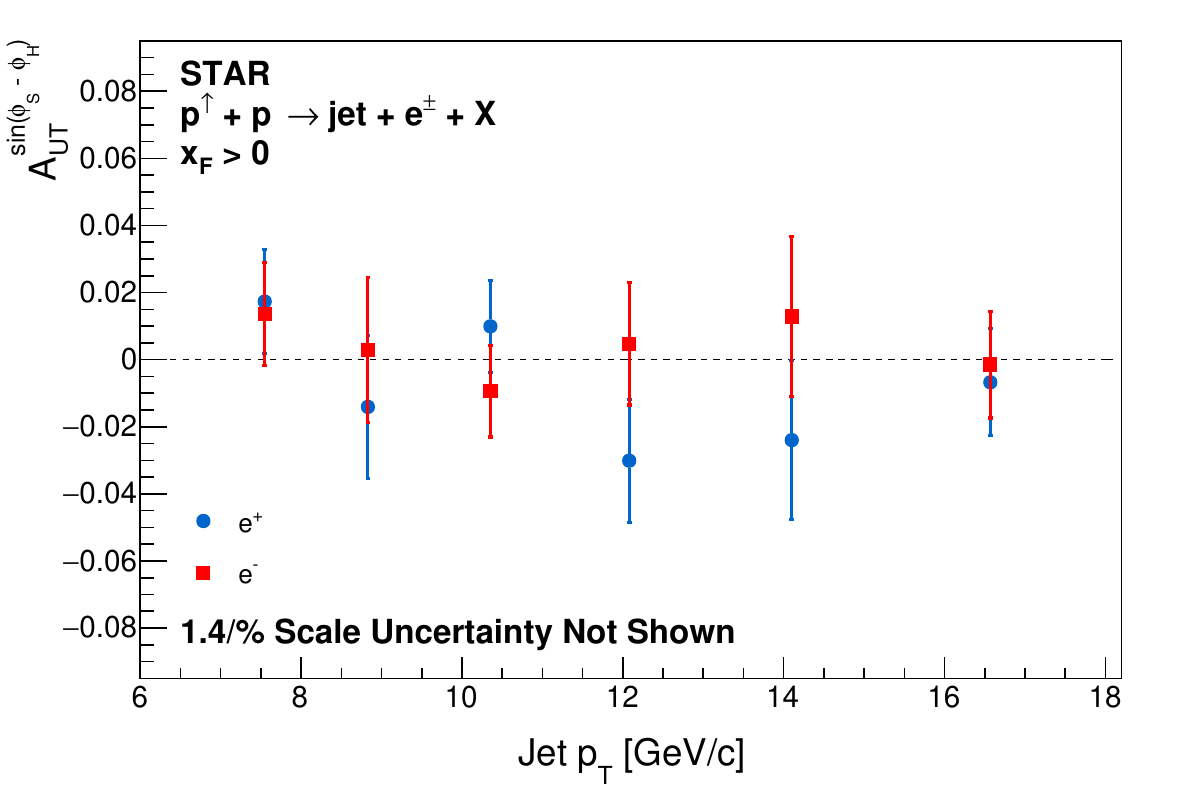}
    \caption{Collins asymmetries, $A_{\mathrm{UT}}^{\sin(\phi_S - \phi_H)}$ for $e^{\pm}$ as a function of jet $p_{\mathrm{T}}$ in $p^{\uparrow}p$ collisions at $\sqrt{s} = 510$ GeV.}
    \label{fig:AUT_e}
\end{figure}

Electrons mainly arise from photon conversions and semi-leptonic decays of heavy-flavor hadrons. The photons primarily originate from $\pi^0$ decay. The asymmetries for $\pi^0$ are expected to be nearly zero, as they are close to the average of the $\pi^+$ and $\pi^-$ asymmetries. Heavy flavor production mainly arises from the $gg \rightarrow q\bar{q}$ process, which does not contribute to the Collins effect since gluon transversity is expected to be zero. To further verify experimentally whether the Collins signal of electrons is indeed consistent with zero, we select a tight electron-enriched region defined by \(n\sigma_{dE/dx} > 4\) and $-5 < n\sigma_{\mathrm{TOF}} < 2$ and then extract Collins asymmetry in this rich region. Fig.~\ref{fig:AUT_e} shows the Collins asymmetry results for electrons and positrons, which, as expected, are consistent with zero within uncertainties.

\FloatBarrier

\mysection{QUANTITATIVE COMPARISON OF $A_{\mathrm{UT}}^{\phi_S-\phi_H}$}

\begin{figure}[H]
    \centering
    \begin{subfigure}[b]{0.48\textwidth}
        \centering
        \includegraphics[width=\linewidth]{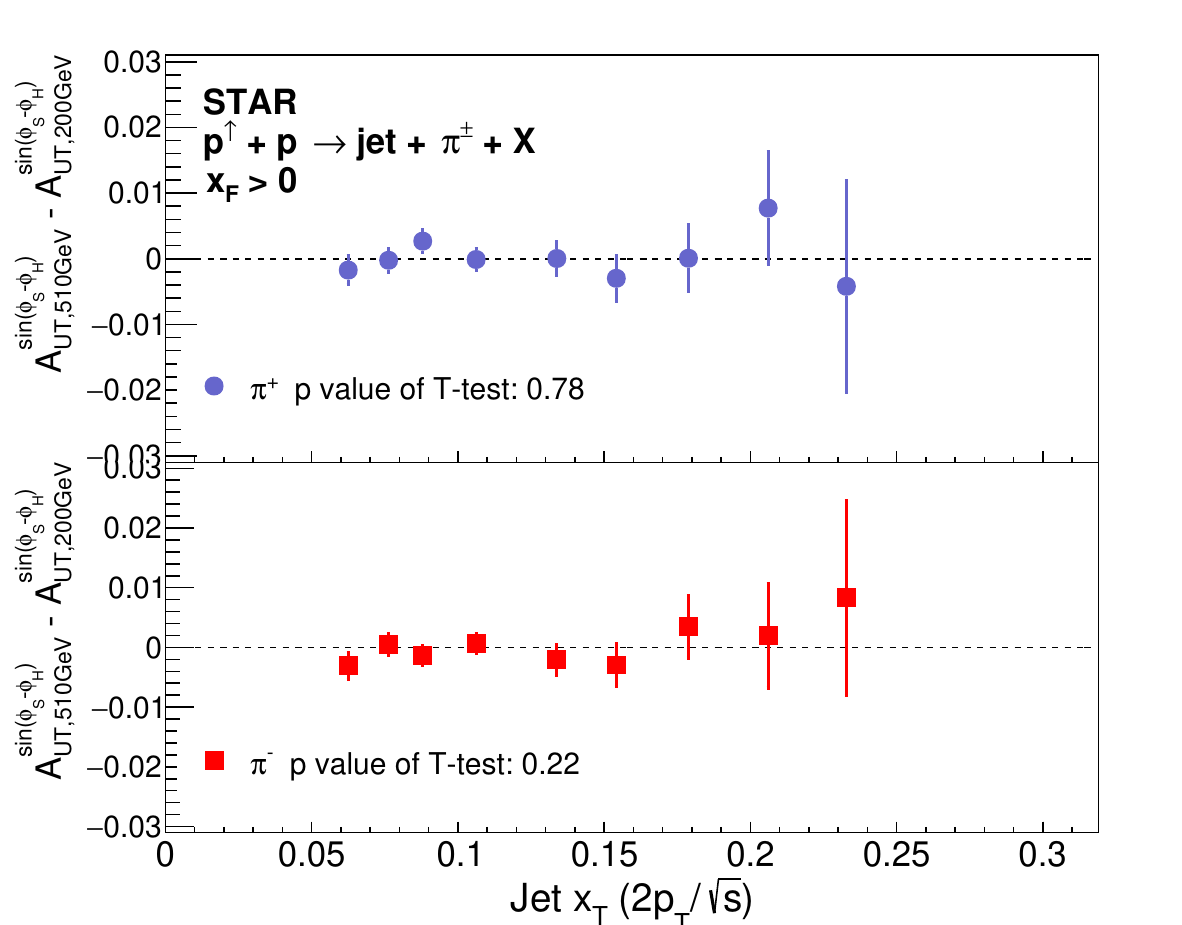}
        \caption{}
        \label{fig1:sub1}
    \end{subfigure}
    \hfill
    \begin{subfigure}[b]{0.48\textwidth}
        \centering
        \includegraphics[width=\linewidth]{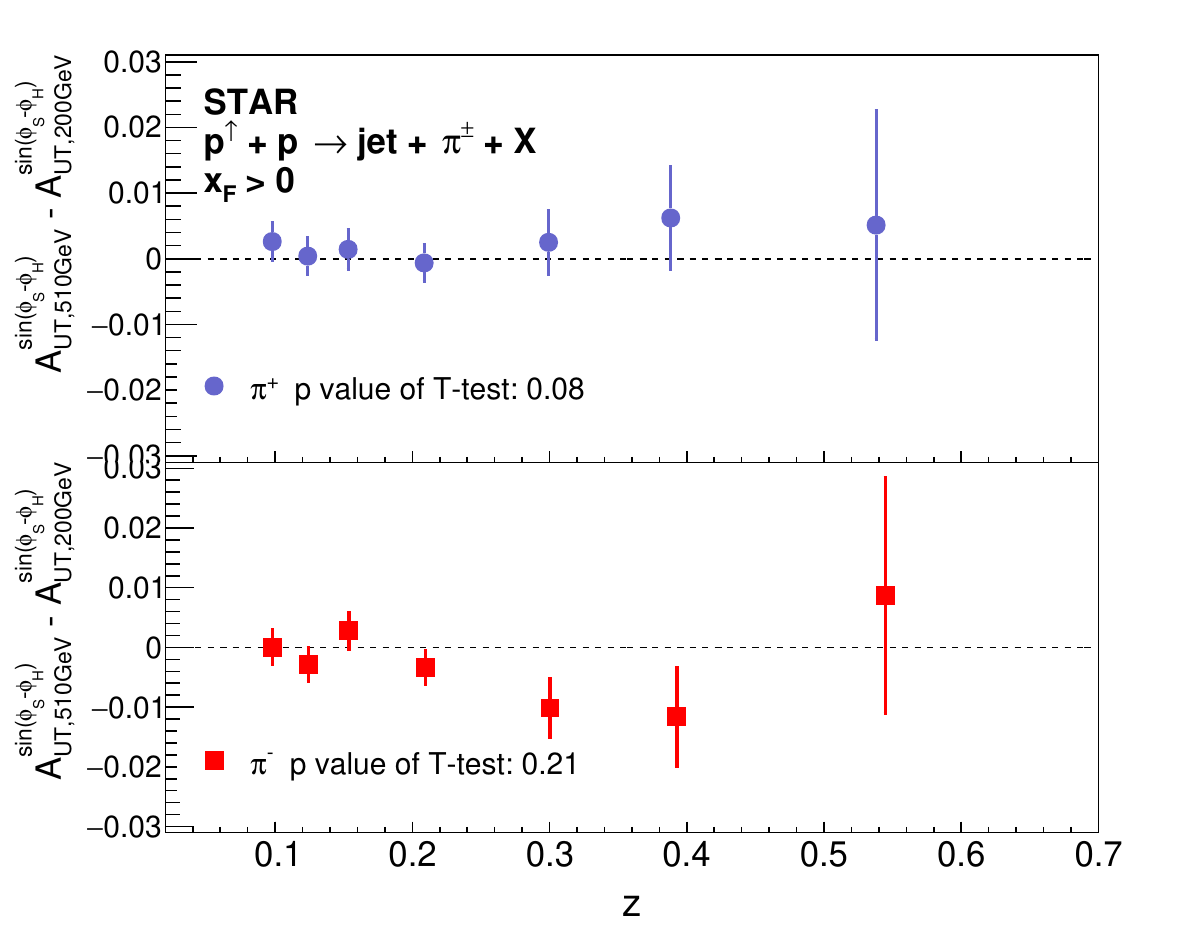}
        \caption{}
        \label{fig1:sub2}
    \end{subfigure}
    \caption{Difference of Collins asymmetries at 510 GeV and 200 GeV, as a function of jet $x_\mathrm{T}$ ($\equiv \frac{2p_\mathrm{T}}{\sqrt{s}}$) (a) and pion's longitudinal momentum fraction $z$ (b) for $\pi^{\pm}$ in $p^{\uparrow}p$ collisions. The error bars encompass the statistical and systematic uncertainties from both the 510 GeV and 200 GeV results, as well as the uncertainties introduced by fitting the 200 GeV data.}
    \label{fig:Diff}
\end{figure}

A $t$-test analysis was conducted to assess the consistency of Collins asymmetries between the two collision energies. To address differences in $x_\mathrm{T}$ and $z$ distributions, the 200 GeV data were adjusted to match the 510 GeV $x_\mathrm{T}$ and $z$ values using a third-order polynomial fit, $i.e.$, $f(x)=ax^3+bx^2+cx$. The shifted 200 GeV results can be expressed as $A_{\mathrm{UT}}^{200 \mathrm{GeV,shift}}=A_{\mathrm{UT}}^{200 \mathrm{GeV}}+f(x^{510 \mathrm{GeV}})-f(x^{200 \mathrm{GeV}})$. To avoid potential correlations between the fitting uncertainties and the data point uncertainties, we removed the point to be shifted in each fitting iteration, performing a separate fit for each point that is shifted. In this case, the uncertainty in the difference between 200 GeV and 510 GeV is composed of three parts: statistical uncertainty, systematic uncertainty, and the uncertainty introduced by the fitting process. Fig.~\ref{fig:Diff}a shows the differences in Collins asymmetries as a function of jet $x_\mathrm{T}$, while Fig.~\ref{fig:Diff}b displays their $z$-dependence under the same $x_\mathrm{T}$ range.  
The differences are statistically consistent with zero, with $p$-values of $0.78$ ($\pi^+$) and $0.22$ ($\pi^-$) for Fig.~\ref{fig:Diff}a, and $0.08$ ($\pi^+$) and $0.21$ ($\pi^-$) for Fig.~\ref{fig:Diff}b – all above the conventional 0.05 significance threshold. These results demonstrate no statistically significant energy dependence of Collins asymmetries within the measured kinematic range.

\FloatBarrier

\mysection{DATA MODEL COMPARISON: $z$ DEPENDENCE OF $A_{\mathrm{UT}}^{\phi_S-\phi_H}$}

\begin{figure}[H]
    \centering
    \includegraphics[width=\textwidth]{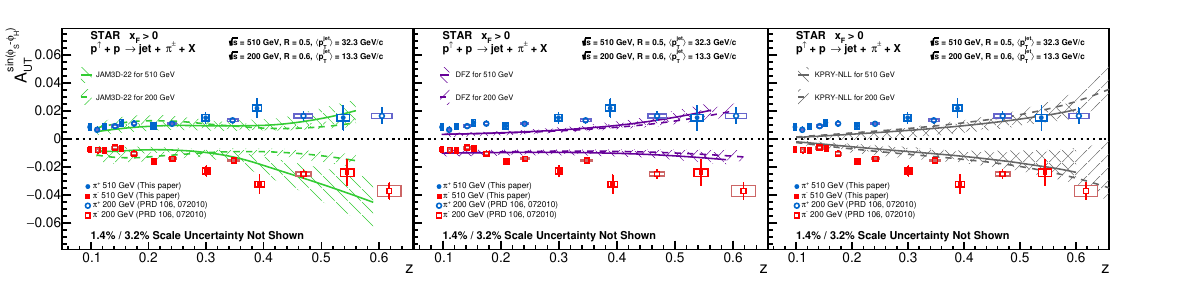}
    \caption{Collins asymmetries, $A_{\mathrm{UT}}^{\sin(\phi_S - \phi_H)}$ as a function of $\pi^{\pm}$ momentum fraction $z$ in $p^{\uparrow}p$ collisions at $\sqrt{s} = 510$ GeV and 200 GeV, together with predictions from three theoretical models: JAM3D-22 (left), DFZ (middle) and KPRY-NLL (right).}
    \label{fig:curves}
\end{figure}

Fig.~\ref{fig:curves} presents a comparison between experimental measurements of the Collins asymmetry as a function of $z$ at both 200 GeV and 510 GeV, and predictions from three theoretical models.

The JAM3D-22 model incorporates both TMD and collinear twist-3 observables, without implementing TMD evolution, and is based on a global fit to data from $e^+e^-$, SIDIS, and $p^{\uparrow}p$ processes~\cite{JAM}. The DFZ model employs the generalized parton model (GPM) at leading order, similarly excluding TMD evolution, and also utilizes inputs from $e^+e^-$, SIDIS, and $p^{\uparrow}p$ data~\cite{DFZ}. The KPRY calculation adopts a hybrid framework that combines collinear and TMD factorization, where the Collins asymmetry arises from the convolution of the collinear transversity distribution with the Collins fragmentation function of the TMD~\cite{KPRY}. This approach includes TMD evolution at next-to-leading logarithmic (NLL) accuracy and is constrained by a global analysis of SIDIS and $e^+e^-$ data.

All three models demonstrate only a weak dependence on the collision energy. Among them, JAM3D-22 shows generally good agreement with the data, whereas both DFZ and KPRY-NLL exhibit minor but noticeable discrepancies in the low-$z$ region.

\FloatBarrier

\mysection{SYSTEMATIC UNCERTAINTIES}

The results in this analysis include three main sources of systematic uncertainty. All estimation methods follow those used in previous STAR Collins analyses~\cite{STAR:2017akg,STAR:2022hqg}.
The dominant one is on the bias introduced by the trigger system during the data collection, as the STAR jet patch trigger is more efficient for certain subprocess than others, for example the relative contribution from quark and gluon initiated jets, and thus bias the measured asymmetries. 
The corresponding systematic uncertainty from trigger bias is estimated using the simulation sample, by calculating the biased-to-unbiased quark and gluon jet fractions, as the Collins asymmetry are sensitive to quark jets only. 
Although this is the dominant source of systematic uncertainty, its magnitude is only about 20\% of the corresponding statistical uncertainty.

The second source is caused by the non-uniform acceptance of the detector, which lead to cross-talk between different azimuthal modulations $i.e.$, Sivers effect and Collins-like effect. 
This is estimated with a data-driven approach, which determine the impact on the Collins modulation by artificially introducing a known asymmetry associated with one physics modulation like Sivers or Collins-like effect.
Its contribution is at the level of approximately 5\% of the statistical uncertainty.

The third source arises from the correction procedure for the Collins angle. The finite resolution of the Collins angle measurement causes a dilution of its distribution, leading to a reduction in the observed asymmetry amplitude. We estimate the Collins angle resolution from simulation data and apply it to a standard $sin(\phi_C)$ function. The corresponding change in the $sin(\phi_C)$ amplitude provides an estimate of the resolution correction factor. By fitting the Collins angle resolution distribution with different functional forms, we obtain slightly different correction factors; their variation is taken as the systematic uncertainty associated with this correction. Overall, this contribution is estimated to be about 2\% of the statistical uncertainty.

As an example, Tab.~\ref{table:pT_results} summarizes the statistical uncertainty and the three sources of systematic uncertainty for the $\pi^+$ Collins asymmetry as a function of jet-$p_T$.

\begin{table}[htb]
    \centering
    \small
    \caption{Collins asymmetry $A_{\mathrm{UT}}$ as a function of jet $p_{\mathrm{T}}$ for $\pi^+$ at $x_{F}>0$. The listed uncertainties correspond to statistical and three systematic sources.}
    \label{table:pT_results}
    \begin{tabular}{cccccc}
        \toprule
        Jet $p_{\mathrm{T}}$ [GeV/$c$] & 
        $A_{\mathrm{UT}}$ & 
        $\delta A_{\mathrm{UT}}^{\mathrm{stat}}$ & 
        $\delta A_{\mathrm{UT}}^{\mathrm{trigger}}$ & 
        $\delta A_{\mathrm{UT}}^{\mathrm{cross}}$ & 
        $\delta A_{\mathrm{UT}}^{\mathrm{reso}}$ \\
        \midrule
        8.57  & -0.00067 & 0.00163 & 0.00001 & 0.00014 & 0.00000 \\
        9.87  & -0.00148 & 0.00193 & 0.00013 & 0.00012 & 0.00000 \\
        12.21 &  0.00177 & 0.00120 & 0.00004 & 0.00010 & 0.00000 \\
        13.98 &  0.00106 & 0.00139 & 0.00004 & 0.00009 & 0.00001 \\
        15.94 & -0.00031 & 0.00142 & 0.00005 & 0.00010 & 0.00000 \\
        19.46 &  0.00411 & 0.00088 & 0.00027 & 0.00006 & 0.00002 \\
        22.44 &  0.00654 & 0.00107 & 0.00041 & 0.00003 & 0.00001 \\
        25.46 &  0.00510 & 0.00104 & 0.00042 & 0.00003 & 0.00002 \\
        29.72 &  0.00817 & 0.00131 & 0.00013 & 0.00002 & 0.00014 \\
        34.13 &  0.01153 & 0.00179 & 0.00002 & 0.00001 & 0.00001 \\
        39.32 &  0.01076 & 0.00264 & 0.00038 & 0.00003 & 0.00004 \\
        45.61 &  0.01790 & 0.00418 & 0.00056 & 0.00007 & 0.00008 \\
        52.55 &  0.02150 & 0.00720 & 0.00067 & 0.00038 & 0.00019 \\
        59.37 &  0.02371 & 0.01211 & 0.00051 & 0.00076 & 0.00004 \\
        \bottomrule
    \end{tabular}
\end{table}

%% file: star-author-list-2025-11-18.listB_aps.txt
\affiliation{Academia Sinica, Nankang, 115, Taipei}
\affiliation{Abilene Christian University, Abilene, Texas   79699}
\affiliation{AGH University of Krakow, FPACS, Cracow 30-059, Poland}
\affiliation{Argonne National Laboratory, Argonne, Illinois 60439}
\affiliation{American University in Cairo, New Cairo 11835, Egypt}
\affiliation{Ball State University, Muncie, Indiana, 47306}
\affiliation{Brookhaven National Laboratory, Upton, New York 11973}
\affiliation{University of Calabria \& INFN-Cosenza, Rende 87036, Italy}
\affiliation{University of California, Berkeley, California 94720}
\affiliation{University of California, Davis, California 95616}
\affiliation{University of California, Los Angeles, California 90095}
\affiliation{University of California, Riverside, California 92521}
\affiliation{Central China Normal University, Wuhan, Hubei 430079 }
\affiliation{University of Illinois at Chicago, Chicago, Illinois 60607}
\affiliation{Chongqing University, Chongqing, 401331}
\affiliation{Creighton University, Omaha, Nebraska 68178}
\affiliation{Czech Technical University in Prague, FNSPE, Prague 115 19, Czech Republic}
\affiliation{Technische Universit\"at Darmstadt, Darmstadt 64289, Germany}
\affiliation{National Institute of Technology Durgapur, Durgapur - 713209, India}
\affiliation{ELTE E\"otv\"os Lor\'and University, Budapest, Hungary H-1117}
\affiliation{Frankfurt Institute for Advanced Studies FIAS, Frankfurt 60438, Germany}
\affiliation{Fudan University, Shanghai, 200433 }
\affiliation{Guangxi Normal University, Guilin, 541004}
\affiliation{University of Heidelberg, Heidelberg 69120, Germany }
\affiliation{University of Houston, Houston, Texas 77204}
\affiliation{Huzhou University, Huzhou, Zhejiang  313000}
\affiliation{Indian Institute of Science Education and Research (IISER), Berhampur 760010 , India}
\affiliation{Indian Institute of Science Education and Research (IISER) Tirupati, Tirupati 517507, India}
\affiliation{Indian Institute Technology, Patna, Bihar 801106, India}
\affiliation{Indiana University, Bloomington, Indiana 47408}
\affiliation{Institute of Modern Physics, Chinese Academy of Sciences, Lanzhou, Gansu 730000 }
\affiliation{University of Jammu, Jammu 180001, India}
\affiliation{Kent State University, Kent, Ohio 44242}
\affiliation{University of Kentucky, Lexington, Kentucky 40506-0055}
\affiliation{Lanzhou University, Lanzhou, 730000}
\affiliation{Lawrence Berkeley National Laboratory, Berkeley, California 94720}
\affiliation{Lehigh University, Bethlehem, Pennsylvania 18015}
\affiliation{Lovely Professional University, Jalandhar - Delhi G.T. Road, Pagwara, Panjab, 144411, India}
\affiliation{Max-Planck-Institut f\"ur Physik, Munich 80805, Germany}
\affiliation{Michigan State University, East Lansing, Michigan 48824}
\affiliation{National Institute of Science Education and Research, HBNI, Jatni 752050, India}
\affiliation{National Cheng Kung University, Tainan 70101 }
\affiliation{Nuclear Physics Institute of the CAS, Rez 250 68, Czech Republic}
\affiliation{The Ohio State University, Columbus, Ohio 43210}
\affiliation{Panjab University, Chandigarh 160014, India}
\affiliation{Purdue University, West Lafayette, Indiana 47907}
\affiliation{Rice University, Houston, Texas 77251}
\affiliation{Rutgers University, Piscataway, New Jersey 08854}
\affiliation{University of Science and Technology of China, Hefei, Anhui 230026}
\affiliation{South China Normal University, Guangzhou, Guangdong 510631}
\affiliation{Sejong University, Seoul, 05006, Korea, Republic Of}
\affiliation{Shandong University, Qingdao, Shandong 266237}
\affiliation{Shanghai Institute of Applied Physics, Chinese Academy of Sciences, Shanghai 201800}
\affiliation{Southern Connecticut State University, New Haven, Connecticut 06515}
\affiliation{State University of New York, Stony Brook, New York 11794}
\affiliation{Instituto de Alta Investigaci\'on, Universidad de Tarapac\'a, Arica 1000000, Chile}
\affiliation{Temple University, Philadelphia, Pennsylvania 19122}
\affiliation{Texas A\&M University, College Station, Texas 77843}
\affiliation{Texas Southern University, Houston, Texas, 77004}
\affiliation{University of Texas, Austin, Texas 78712}
\affiliation{Tsinghua University, Beijing 100084}
\affiliation{University of Tsukuba, Tsukuba, Ibaraki 305-8571, Japan}
\affiliation{University of Chinese Academy of Sciences, Beijing, 101408}
\affiliation{United States Naval Academy, Annapolis, Maryland 21402}
\affiliation{Valparaiso University, Valparaiso, Indiana 46383}
\affiliation{Variable Energy Cyclotron Centre, Kolkata 700064, India}
\affiliation{Warsaw University of Technology, Warsaw 00-661, Poland}
\affiliation{Wayne State University, Detroit, Michigan 48201}
\affiliation{Wuhan University of Science and Technology, Wuhan, Hubei 430065}
\affiliation{Yale University, New Haven, Connecticut 06520}

\author{B.~E.~Aboona}\affiliation{Texas A\&M University, College Station, Texas 77843}
\author{J.~Adam}\affiliation{Czech Technical University in Prague, FNSPE, Prague 115 19, Czech Republic}
\author{L.~Adamczyk}\affiliation{AGH University of Krakow, FPACS, Cracow 30-059, Poland}
\author{I.~Aggarwal}\affiliation{Panjab University, Chandigarh 160014, India}
\author{M.~M.~Aggarwal}\affiliation{Panjab University, Chandigarh 160014, India}
\author{Z.~Ahammed}\affiliation{Variable Energy Cyclotron Centre, Kolkata 700064, India}
\author{A.~K.~Alshammri}\affiliation{Kent State University, Kent, Ohio 44242}
\author{E.~C.~Aschenauer}\affiliation{Brookhaven National Laboratory, Upton, New York 11973}
\author{S.~Aslam}\affiliation{Fudan University, Shanghai, 200433 }
\author{J.~Atchison}\affiliation{Abilene Christian University, Abilene, Texas   79699}
% ...

\collaboration{STAR Collaboration}\noaffiliation

%% file: main.bbl
%